\documentclass[prl,twocolumn,showpacs,preprintnumbers,amsmath,amssymb,floats]{revtex4}

\usepackage{graphicx}  
\usepackage{subeqnar}  

\begin{document}
\draft \preprint{}

\title{Zooming-in on the charge ordering in YBa$_2$Cu$_3$O$_{6.5}$}

\author{D. L. Feng$^{1,3}$, A. Rusydi$^2$, P. Abbamonte$^4$, L. Venema$^2$, I. Elfimov$^{1}$,
R. Liang$^{1}$, D. A. Bonn$^{1}$, W. N. Hardy$^{1}$, C.
Sch\"{u}\ss ler-Langeheine$^{5}$, S. Hulbert$^4$, C.-C. Kao$^4$
and G. A. Sawatzky$^{1,2}$ }

\affiliation{$^1$Department of Physics and Astronomy, University
of British Columbia, Vancouver, BC, V6T 1Z1, Canada.}
\affiliation{$^2$Materials Science Centre, University of
Groningen, 9747 AG Groningen, The Netherlands.}
\affiliation{$^3$Department of Physics, Applied Surface Physics
State Key Laboratory, and Synchrotron Radiation Research Center,
Fudan University, Shanghai 200433, China}
\affiliation{$^4$National Synchrotron Light Source, Brookhaven
National Laboratory, Upton, NY 11973, USA.}
 \affiliation{$^5$ II.
Physikalisches Institut, Universit$\ddot{a}$t zu K\"{o}ln, D-50937
K\"{o}ln, Germany}

\date{\today}

\begin{abstract}
We report direct evidence of charge/orbital ordering of low energy
electronic states of $Cu$ in YBa$_2$Cu$_3$O$_{6+x}$ ortho-II phase
in both the $CuO_3$ chain and the CuO$_2$ plane. Huge enhancement
of the $(\frac{1}{2},0,0)$ superstructure Bragg peak is observed
when photon energy is tuned to the $Cu\,L_{2,3}$ absorption edge
with large polarization dependence. The ordering in the $CuO_2$
plane discovered here sheds new light on how the one dimensional
$Cu-O$ chains affect the $CuO_2$ plane, and why many normal and
superconducting state properties of this system exhibit strong
anisotropy.
\end{abstract}


\maketitle

YBa$_2$Cu$_3$O$_{6+x}$ is one of the most studied high temperature
cuprate superconductors. Many important observations, such as the
neutron resonance mode\cite{Mook93} and d-wave superfluid
behavior,\cite{Hardy93} were first made on this system. However,
whether the detailed properties of these phenomena are generic to
all cuprates still remains questionable due to the presence of
$CuO_3$ chains in addition to $CuO_2$ planes in this material. So
far, virtually all known experimental methods have been used to
study this system, but little is known about the role that $CuO_3$
chains play in the low energy electronic structure, or if the
chain is superconducting or even conducting at all. For example,
surface reconstruction of YBa$_2$Cu$_3$O$_{6+x}$ prevents angle
resolved photoemission spectroscopy or scanning tunnelling
spectroscopy from obtaining reliable electronic structure
information.

The Ortho-II chain-ordered phase of YBa$_2$Cu$_3$O$_{6.5}$ is an
ideal system for studying the local interplay between the chain
and the plane. Conventional x-ray scattering experiments on this
system exhibit a superlattice Bragg peak corresponding to a
doubling of the unit cell. The doubling is modelled with oxygen
atoms in the $CuO_3$ chain layers being ordered into alternating
full and empty chains, as illustrated in Fig. 1a.
\cite{Andersen99,Zimmermann99,Liang00} If the chain does affect
the plane, one would expect charge ordering in the plane induced
by this chain ordering, and by studying its nature, one can study
how the electronic structure and the superconductivity in the
$CuO_2$ plane are affected. However, in practice, this is a
difficult task because besides the primary order, i.e. alternate
filling of oxygen accompanied by alternation in the valence states
of chain $Cu$'s, there are additional structural orders of the $Y$
and $Ba$ heavy ions. These ions are located between chains, and
chain ordering changes their positions. Since all the electrons in
the system contribute to conventional x-ray scattering with large
contributions from heavy atoms, the primary order is not directly
studied in such experiments. That the chain ordering causes these
changes in position is a plausible speculation, but the proposed
model of oxygen ordering in YBa$_2$Cu$_3$O$_{6.5}$ presented in
Fig. 1a and especially the accompanying change in $Cu$ oxidation
states remains to be verified through more direct experiments.

We applied resonant soft x-ray scattering to this problem, which
is a novel diffraction technique that can provide element specific
and spatial information of the valence electron
distribution.\cite{Blume94,Abbamonte02} We found that the valence
state of the chain $Cu$ is ordered, moreover, this ordering
induces significant charge and orbital ordering in the plane. This
1D footprint of the chain on the plane gives a plausible
explanation for many of the observed unusual phenomena in
$YBa_2Cu_3O_{6+x}$.

\begin{figure}[t!]

\centerline{\includegraphics[width=3.3in]{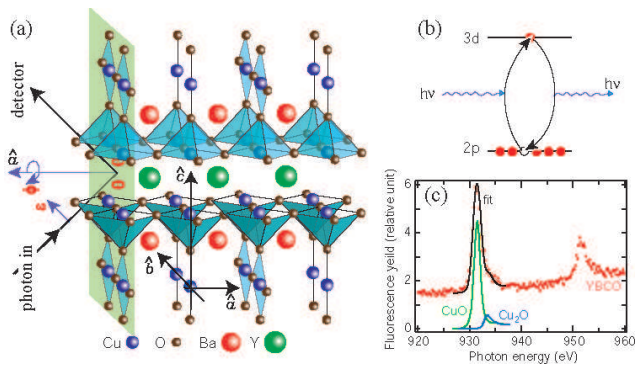}}
\caption{The system understudy and the experimental setup. (a)
illustrates the proposed ortho-II ordered phase of
YBa$_2$Cu$_3$O$_{6.5}$. The chain filled with oxygen (full chain)
and the one without oxygen (empty chain) correspond to $Cu$
oxidation states $Cu^{2+}$ and $Cu^{1+}$ respectively. The $
\theta/2\theta$ Bragg scattering experimental geometry is also
shown, with photon polarization $\varepsilon$  in the $
\theta/2\theta $ plane. The sample can be rotated azimuthally
around the crystalline $\hat{a}$-axis without breaking the Bragg
condition. The azimuthal angle is defined to be $0^{\circ}$ in the
geometry shown, where the $ \theta/2\theta$ plane and the $ac$
plane coincide. (b) The $Cu\,2p\!\rightarrow\!3d$ resonant
scattering process which dramatically enhances the scattering from
$Cu\,3d$ orbitals. (c) Fluorescence yield XAS spectrum for
$Cu\,2p$ to $3d$ transitions in YBa$_2$Cu$_3$O$_{6.5}$, which can
be fitted by a linear combination of $CuO$ and $Cu_2O$ XAS spectra
taken from ref.\cite{Tjeng92}. } \label{autonum1}

\end{figure}

The  Resonant soft x-ray scattering (RSXS) experiments were
performed on untwinned ortho-II phase YBa$_2$Cu$_3$O$_{6.5}$
single crystal at the soft x-ray Beamline X1-B of National
Synchrotron Light Source (NSLS), using the ``Spinoza
diffractometer" in an ultra-high-vacuum (UHV) chamber with a base
pressure of $2\times10^{-10} mbar$ as described in
ref.\cite{Abbamonte02}. UHV is needed because of the surface
sensitivity of the technique, as the x-ray penetration depth in
this experiment is about 1200$\AA$. The sample was prepared as
described in ref.\cite{Liang00} with a superconducting transition
temperature of 62K. Data were collected at 100K.

As illustrated in Fig. 1b, when the photon energy is tuned to the
$Cu\,2p\!\rightarrow\!3d$ transition, the contribution from $Cu$
atoms to the scattering is greatly enhanced through a virtual
excitation process called resonant or anomalous
scattering\cite{Blume94}. In the dipole approximation, it is
described by the atomic form factor:

\begin{eqnarray}
\nonumber f \propto \sum_f |\sum_m  \frac{\langle \psi_{f},
\omega_1 \! \mid \!\mathbf{\varepsilon}\! \cdot \!\mathbf{r}\!
\mid \psi_{m} \rangle \langle \psi_{m} \! \mid \!
\mathbf{\varepsilon'}\! \cdot \!\mathbf{r}\! \mid \!
\psi_{i},\omega_0 \rangle
 }{E_m-E_i-\hbar\omega_0-i\Gamma_m} | \\
\nonumber \cdot \delta (E_f+\hbar\omega_1-E_i-\hbar\omega_0)
\end{eqnarray}

\noindent where $ \psi_{i}$, $ \psi_{m}$, and $\psi_{f} $ are the
initial, intermediate, and final state wavefunctions with energies
of $E_i$, $E_m$, and $E_f$ respectively. $ \Gamma_m $ is the life
time energy broadening of the intermediate state. $\omega_0$ and
$\omega_1$ are the incoming and scattered photons with
polarization $\varepsilon$ and $ \varepsilon'$ respectively.
Apparently, the scattering signal contains information of
electronic structure of the states involved. In this case, if $
\psi_{i}$ contains $3d$ holes, $Cu\,2p\!\rightarrow\!3d$
transition is allowed to the state $ \psi_{m}$, and therefore the
$Cu\,3d$ states and their spatial distribution are probed
directly. On the contrary, ordinary resonant hard x-ray scattering
based on $Cu\,1s\!\rightarrow\!4p$ transition only probes empty
$4p$ states, which contribute only weakly to the low energy
physics in the problem. The $Cu\,2p\!\rightarrow\!3d$ transition
has been extensively studied by x-ray absorption spectroscopy
(XAS), where two peaks around 930 eV and 950 eV correspond to
final states with $2p_{3/2}$ and $2p_{1/2}$ core holes
respectively, referred to as the $Cu$ $L_3$ and $L_2$ absorption
edges (Fig. 1c). The YBa$_2$Cu$_3$O$_{6.5}$ $L_3$ edge can be
fitted by a linear combination of $CuO$ and $Cu_2O$ spectra,
consistent with the model in Fig. 1a, where $Cu$ is in a nominally
$Cu^{2+}$($3d^9$) in the full chain, and $Cu^{1+}$($3d^{10}$)
state in the empty chain. Due to hybridization with oxygen, there
are still some $3d$ holes on $Cu^{1+}$, contributing weakly to
$2p\!\rightarrow\!3d$ transition in YBa$_2$Cu$_3$O$_{6.5}$ at an
energy $\sim$2eV higher than the main $Cu^{2+}$ peak. Because the
resonant scattering arises via the dipole allowed
($2p^63d^9\!\rightarrow\!2p^53d^{10}$) transitions, it is
dominated by $Cu$ atoms in the empty or full chains at different
photon energies, resulting in a large contrast and a huge
enhancement of the superlattice Bragg peak.

We will now use this contrast in the $Cu^{1+}$ and $Cu^{2+}$
scattering to study the ordering of $Cu\,3d$ holes. Data were
taken in a $\theta/2\theta$ Bragg diffraction geometry as
illustrated in Fig. 1a. The scattering intensity as a function of
angle and photon energy, or resonance profile, is shown in Fig.
2a. The (1/2,0,0) Bragg peak corresponding to the Ortho-II
superstructure is observed at all photon energies, with two
resonances occurring near the $L_{2,3}$ absorption edges. The
intensity far from the Bragg angle is very weak, which indicates
that the contamination to the Bragg peak from isotropic diffuse
scattering and fluorescence is negligible. Compared with the
off-resonance intensity at 920 eV, the measured Bragg peak
intensity is enhanced by a factor of 62 at the $L_3$ edge, and 8
at the $L_2$ edge. The weak enhancement at the $L_2$ is mainly
caused by the competing non-radiative Coster-Kronig decay channel
of the $2p_{1/2}$ core-hole.\cite{Antonides77}

For the non-resonant scattering, the ordered $O^{2-}$ ions in
chains contribute on the order of $10^2=100$, while the
alternating chain $Cu^{2+}$ and $Cu^{1+}$ ions just contribute on
the order of $1^2=1$. As there is additional scattering from the
$Y^{3+}$ and $Ba^{2+}$ ions, the $Cu$ contribution to the
off-resonance scattering is estimated to be less than 1\%.
Therefore, the $Cu$ contribution is enhanced by more than 6200
times at the $L_3$ edge. Such a gigantic enhancement is rarely
observed, and far beyond what one would expect if $Cu$ ions are
simply displaced from their stoichiometric position with no change
in valence. [Actually, $Cu$ ions are situated in between either
two full or two empty chains, its displacement (if any) should be
very small.] Therefore, this observation directly proves that the
local electronic structure or electron density of the $Cu\,3d$
orbitals is modulated with a period of $2a$ perpendicular to the
chain in the Ortho-II phase, which should have a significant
effect on the low energy electronic structure.

\begin{figure}[t!]
\vspace*{-0.5cm}\hspace*{0cm}
\centerline{\includegraphics[width=3.3in]{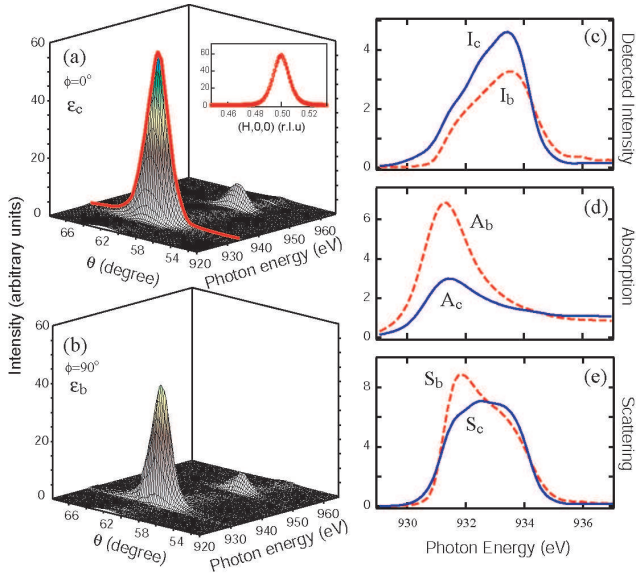}}
\caption{Resonant soft x-ray scattering results. (a) and (b), The
scattered photon intensity as a function of energy and angle for
$\phi=0^{\circ}$ and $90^{\circ}$ respectively. Inset: The
(1/2,0,0) Bragg peak at resonance (i.e. the thick curve through
the $L_3$ peak in a) in reciprocal lattice unit, which is well
described by an anisotropic Lorentzian
function,\cite{Andersen99,Liang00} giving a coherence length of
about 60\AA. (c) The integrated (over momentum) Bragg intensity;
(d) the x-ray absorption; and (e), the absorption-corrected
scattering as a function of the photon energy.Arbitrary units are
used here.} \label{autonum2}
\end{figure}

Having determined that the $Cu\,3d$ holes are ordered, the next
step is to zoom-in even further and try to distinguish the
ordering of the chain $Cu$ from that of the plane Cu. Because the
holes are mainly situated in the $ab$ plane for the plane Cu, and
$bc$ plane for the chain Cu, their contribution to the scattering
cross-section will have very different polarization dependencies.
Therefore, by comparing scattering intensities taken at different
polarizations, one can differentiate between chain and plane
$Cu\,3d$ hole orbitals, and separately investigate their ordering.
The polarization can be changed by using different azimuthal
angles $\phi$  (Fig. 1a).  For $\phi=0^{\circ}$, and $90^{\circ}$,
polarization $\varepsilon_c =
\cos(\theta_B)\hat{a}+\sin(\theta_B)\hat{c}$ and $\varepsilon_b =
\cos(\theta_B)\hat{a}+\sin(\theta_B)\hat{b}$ respectively (denoted
by subscript $c$ and $b$ hereafter as the $\hat{a}$-component is
fixed), where $\theta_B$ is the Bragg angle. Figs. 2a and 2b are
the resonance profiles for the $L_{2,3}$ edge at $\phi=0^{\circ}$
and $90^{\circ}$ respectively. They are normalized at the
non-resonant intensity on the low photon energy side, where the
atomic form factor is basically independent of the polarization.
The momentum-integrated Bragg peak intensity ($I$) does show
significant polarization dependence as shown in Fig. 2c for the
$L_3$ edge. As the incoming and scattered photons will be absorbed
in the material, the measured data (Fig. 2c) has to be corrected
for the polarization dependent self-absorption effect by $I=S/A$
where $S$ and $A$ are the scattering and absorption cross-sections
respectively. \cite{Als01} The x-ray absorption in
YBa$_2$Cu$_3$O$_{6.5}$ has been well studied, and is shown in Fig.
2d for the two azimuthal angles.\cite{Nucker95,Merz98} The
absorption-corrected resonance scattering as a function of photon
energy is shown in Fig. 2e, which gives the ratio of the
integrated total scattering weight over the entire $L_3$ edge for
two azimuthal angles is $\int\! S_{c}dE/\int\!
S_{b}dE=0.92\pm0.05$. One finds that scattering and absorption
have different lineshapes here, and resonance of the scattering
arises at slightly higher photon energies (about 0.5 eV) than the
absorption. This is because the scattering is governed by the
amplitude and absorption by the imaginary part of the atomic form
factor.\cite{Als01}

The measured weaker scattering for light partially polarized along
$\hat{c}$ than along $\hat{b}$ cannot be understood if one assumes
a homogeneous plane. As illustrated in Fig. 3a, the $CuO_3$
plaquet in the full chain is squeezed by about 4.7\% in the
$\hat{c}$ direction. Both DFT calculations and quantum chemistry
analysis show that this would cause a mixing of the chain
$Cu\,3d_{3x^2-r^2}$ and $3d_{y^2-z^2}$ orbitals and relocate more
holes along the $\hat{c}$ axis. In the empty chain (Fig. 3b), the
$O-Cu-O$ dumbbell structure also drives a hole into the
$Cu\,3d_{3z^2-r^2}$ orbital, so that most of the holes are
situated along the $\hat{c}$ direction. Because the energy
distribution of holes is centered at different positions for
different chains as suggested by XAS data (Fig.1b),  both the full
and empty chain scatter $\hat{c}$-polarized light more than the
$\hat{b}$-polarized light, opposite to the observation. This
clearly shows that substantial charge density modulations also
exist in the $CuO_2$ plane layer to give more scattering to the
$\hat{b}$-polarized light.

\begin{figure}[t!]
\vspace*{-0.5cm}\hspace*{0cm}
\centerline{\includegraphics[width=3.3in]{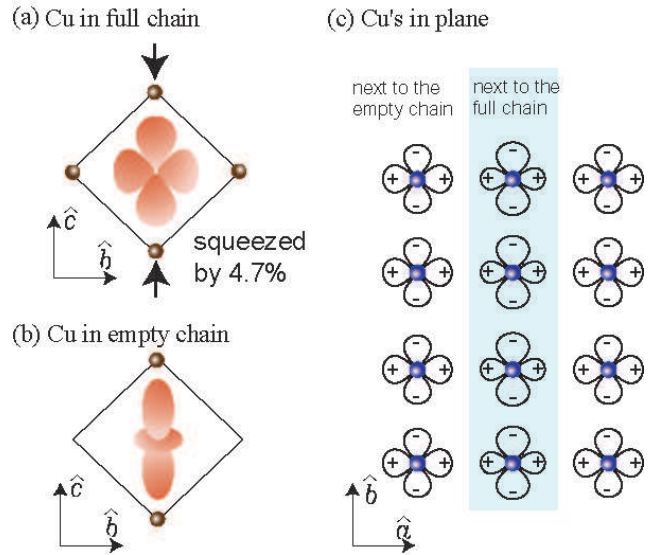}}
 \caption { Estimation of the Charge and orbital ordering in the CuO$_2$
plane. Cartoons of the hole wave functions in the (a) full chain
and (b) empty chain. (c) is a cartoon of possible charge/orbital
ordering of the holes in different planar $Cu$ ions that can
reproduce the data.} \label{autonum3}
\end{figure}

Besides the qualitative claims that one can derive from RSXS, it
is even more important that it could provide (if not pin down)
some quantitative information on the local electronic structure.
In the case here, the measured $S_{c}$ and $S_{b}$ directly relate
to the occupation of the $e_g$ orbitals of the four different
types of $Cu$ ions in the unit cell. Although it is not possible
to determine these eight functions of energy based on $S_{c}$ and
$S_{b}$, in principle, a full determination should be possible,
after further information is obtained with more experiments
performed in cleaner $\sigma\sigma$ experimental geometry and at
more azimuthal angles. we leave this for future studies.

This planar charge/orbital ordering is associated with many
factors, including the Madelung potential of extra O ions in the
full chain which attract more holes in the planar $Cu$ underneath,
the long range Coulomb interaction between holes, the
hybridization between the $p_z$ orbital of the apical $O$ (shared
by the chain and plane) and the planar $Cu\,3d_{3z^2-r^2}$ orbital
and so on. Our simple quantum chemical estimations with parameters
established in experiments and numerical studies,\cite{Eskes92}
indicates that to fit the data, the charge modulation is on the
order of 0.05 holes in planar $Cu$ ions and accompanied by orbital
modulations, \textit{i.e.} different mixing of planar
$3d_{x^2-y^2}$ and $3d_{3z^2-r^2}$ orbitals at different sites as
illustrated by a cartoon in Fig. 3c. Recent nuclear quadruple
resonance (NQR) experiments observed four different $Cu$ sites
corresponding to two planar and two chain $Cu$ sites.\cite{Yamani}
We believe it is directly related to the charge/orbital ordering
discussed here, as the electric field gradient at the $Cu$ nucleus
is very sensitive to the relative occupation of $3d_{x^2-y^2}$ and
$3d_{3z^2-r^2}$ orbitals since these two orbitals contribute with
opposite signs to the electric field gradient.

We have shown that even if the chains are insulating, as most
defected 1D system would be, they imprint their 1D nature onto the
planar low lying electronic states. This charge/orbital ordering
would cause noticeable effects on the superconductivity because of
the high sensitivity to doping of the cuprates. Without having to
include stripes as a separate physical effect,\cite{Tranquada95}
the planar charge/orbital ordering discussed here might provide an
alternative plausible explanation for the observed extraordinarily
strong anisotropy in its superfluid density,\cite{Basov95}
superconducting gap,\cite{Lu01} and conductivity,\cite{Ando02}
which are all significantly larger for the $\hat{b}$ direction
than for the $\hat{a}$ direction. It might also provides a
preferential direction for the observed incommensurate magnetic
fluctuation along $\hat{a}$, although we note that this magnetic
fluctuation has a much larger length scale
($\sim10a$).\cite{Mook00,Dai01}

Strongly correlated systems are known to exhibit rich structural,
electronic and magnetic order, which are driven by only a few
valence electrons.\cite{Masatoshi98,Tokura00} Extensive and
detailed neutron and magnetic x-ray studies have revealed the
magnetic correlations and
order.\cite{Mook00,Dai01,Kastner98,Gibbs99} In this paper we
present the first direct studies in cuprates of what happens in
the charge channel, demonstrating strong charge modulations
distinct from the ordering observed in the spin channel. The
spatial information of the electronic states revealed partially
here sets the stage for understanding of many anomalous properties
in the system. We expect more comprehensive understanding of many
other strongly correlated systems will be reached with the help of
RSXS.

We gratefully ackowledge the helpful discussions with Dr. John
Tranquada and Prof. Changyoung Kim, and help from Dr. David Broun
and Mr. Patrick Turner with the sample preparations. This work is
supported by the Natural Science and Engineering Research Council
of Canada, NWO (Dutch Science Foundation) via the Spinoza program
who funded the ``Spinoza diffractometer", and Netherlands
Organization for Fundamental Research on Matter (FOM). Research
carried out at the NSLS is supported by the U.S. DOE. DLF is also
supported by the Canadian Institute for Advanced Research (CIAR)
and National Science Foundation of China.

\end{document}